\documentclass[sigconf]{acmart}
\usepackage{subfigure}

\AtBeginDocument{%
  \providecommand\BibTeX{{%
    \normalfont B\kern-0.5em{\scshape i\kern-0.25em b}\kern-0.8em\TeX}}}

\setcopyright{rightsretained}
\acmConference[EARS '20]{ACM SIGIR International Workshop on Explainable Recommendation and Search}{July 30, 2020}{Virtual Event, China}
\acmBooktitle{EARS '20: ACM SIGIR International Workshop on Explainable Recommendation and Search,
July 30, 2020, Virtual Event, China}
\acmISBN{978-1-4503-8016-4/20/07}
\acmPrice{15.00}
\copyrightyear{2020}
\acmYear{2020}
\acmDOI{}

\begin{document}

\title{Helping results assessment by adding explainable elements to the deep relevance matching model}

\author{Ioannis Chios and Suzan Verberne}
\email{s.verberne@liacs.leidenuniv.nl}
\affiliation{%
  \institution{Leiden Institute of Advanced Computer Science, Leiden University}
}


\begin{abstract}
In this paper we address the explainability of web search engines. We propose two explainable elements on the search engine result page: a visualization of query term weights and a visualization of passage relevance.
The idea is that search engines that indicate to the user why results are retrieved are valued higher by users and gain user trust. 
We deduce the query term weights from the term gating network in the Deep Relevance Matching Model (DRMM) and visualize them as a doughnut chart. In addition, we train a passage-level ranker with DRMM that selects the most relevant passage from each document and shows it as snippet on the result page. Next to the snippet we show a document thumbnail with this passage highlighted. We evaluate the proposed interface in an online user study, asking users to judge the explainability and assessability of the interface.
We found that users judge our proposed interface significantly more explainable and easier to assess than a regular search engine result page. However, they are not significantly better in selecting the relevant documents from the top-5. 
This indicates that the explainability of the search engine result page leads to a better user experience. Thus, we conclude that the proposed explainable elements are promising as visualization for search engine users. 
\end{abstract}

\keywords{Explainable search, Query analysis, Passage ranking, User interface, User evaluation}

\maketitle

\section{Introduction}
Search engines rely on advanced machine learning  models for document ranking. For users it can be difficult to understand why the retrieved documents are relevant to their query. This has motivated a line of research to make search engines and recommender systems explainable to their users  \cite{zhang2019sigir}. The idea is that if the system indicates to the user why results are retrieved or recommended, it gains trust from the user. There has been more attention to explainable recommendation than to explainable search, with a focus on explaining the user profiling aspect of recommender systems \cite{zhang2019ears}. In this paper, we address the explainability of non-personalized web search engines. We focus on the explanation of query--document relevance on the search engine result page.

The most common type of explanation of relevance on the result page is the search snippet, giving a preview of each retrieved document and thereby an indication of its relevance. Query keywords are typically marked in boldface in the snippet to show the user the query relevance of the document. The interpretability of search result snippets has been investigated by Mi and Jiang \cite{mi2019understanding}. They found that snippets play an important role in explaining why documents are retrieved and how useful those documents are.


In this paper we follow-up on the work by Mi and Jiang by further investigating the explainability potentials of neural ranking models and  proposing a visualization of the explainable model aspects on the search engine result page. 

We attack the explainability task on two levels: on the query level and on the document level. We make the importance of each query term explicit using features of the model's architecture and training process. Additionally, we split the document in smaller passages to investigate the different matching scores between the query and individual passages. We propose an explainable user interface that shows the query term relevance as a doughnut chart, and the passage relevance as a document thumbnail with passage highlights. We evaluate our explainable interface in a small-scale user study in which participants judge the search engine result page on explainability and assessability. 

We address the following research questions: \begin{enumerate} \item What is the ranking effectiveness of DRMM when selecting the most relevant passage of each document? \item How do users judge the explainability and assessability of our explainable search engine result page compared to a regular result page? \item How well can users select the relevant documents based on only the snippets on the result page, in the explainable interface compared to the regular interface? \end{enumerate}

In the remainder of this paper, we first discuss the background research on neural information retrieval and explainable search (Section~\ref{sec:bg}), then we describe our methods (Section~\ref{sec:methods}) and retrieval experiments (Section~\ref{sec:experiments}). In Section~\ref{sec:userstudy} we describe the design and results of the user study for evaluating our explainable interface. We conclude our paper in Section~\ref{sec:conclusion} with answers to the research questions and suggestions for future work.\footnote{Our code is available at \url{https://github.com/giannisosx/explainable-search-drmm/}}

\section{Background} \label{sec:bg}
\subsection{Neural information retrieval}
Neural ad hoc retrieval is typically approached as a two-step process: a retrieval step followed by a re-ranking step. The purpose of this combination is the reduction of the computational demands of the neural ranking models. In the first step, K candidate documents for each query are retrieved using a traditional IR baseline. Then the retrieved documents are re-ranked using the neural model in order to achieve better performance in limited training time. 
This is a procedure widely adopted in prior work on neural ranking for ad-hoc retrieval \cite{guo2016deep,mcdonald2018deep,macavaney2019cedr}. In this paper we build our explainability methods on the Deep Relevance Matching Model (DRMM)~\cite{guo2016deep}.

Over the last years, various IR toolkits have been built by the scientific community to facilitate the development of novel information retrieval methods. Anserini~\cite{yang2017anserini} is an open source information retrieval toolkit built on top of Lucene.\footnote{\url{http://anserini.io}, \url{https://lucene.apache.org/}} 
The advantage of Anserini over other research oriented IR toolkits such as Indri and Galago is its efficiency in indexing and retrieval of large document collections. 

A commonly used framework for neural re-ranking is MatchZoo~\cite{fan2017matchzoo}, an open-source framework that implements various state-of-the-art models for neural ranking. MatchZoo facilitates the design, the comparison and the sharing of deep text matching models.\footnote{\url{https://github.com/NTMC-Community/MatchZoo}} It offers implementations for data processing, neural matching models as well as modules for training and evaluation. Its data processing module contains standard text preprocessing methods such as word tokenization, stemming etc. using NLTK.\footnote{\url{https://www.nltk.org/}}


\subsection{Explainable search}






Recent work on explainable search has been focused on three directions: explainable product search \cite{ai2019explainable}, interpretability of neural retrieval models \cite{singh2019exs,fernando2019study} and their axiomatic analysis \cite{rennings2019axiomatic}. Explainable product search is closely related to explainable recommendation since both aim at enhancing the user experience in online shopping and hence in increasing the profits of e-commerce companies. Axiomatic analyses of neural retrieval models investigate to what extent the formal constraints \cite{fang2004formal} of retrieval models are satisfied from neural ranking models. 

On the other hand, research on the interpretability of neural retrieval models attempts to give more insights to the user regarding the search process. Although current search engines have made steps towards the transparency of their results, many users still have little understanding on how they work.  Singh and Anand \cite{singh2019exs} developed an explainable search engine designed to aid the users in the search task. They used a modified version of LIME \cite{ribeiro2016model}, adapted for rank learning, to interpret the results of neural ranking algorithms. Moreover, they designed an interface for their search engine in which they visualized the explanations. 

The work by Singh and Anand does not involve user feedback for evaluating how helpful the search engine is according to the users. Mi and Jiang \cite{mi2019understanding} conducted an extensive user study to address explainability of result pages. They evaluated a commercial search engine in which the document snippets are the explainable elements on the result page. In the current paper, we build on this work by implementing an explainable neural search engine and evaluating its retrieval effectiveness as well as its explainable elements with user feedback.

\section{Methods}\label{sec:methods}
Our methods comprise three parts: the general architecture of retrieval and re-ranking (Section~\ref{sec:retrieval}), neural re-ranking (Section~\ref{sec:reranking}), and the implementation of explainability elements (Section~\ref{sec:explainability}).

\subsection{Retrieval architecture} \label{sec:retrieval}
We use Anserini to build an inverted index for our document collection. For each query we retrieve the top K documents with BM25 \cite{BM25robertson1994some}. We use the relevance assessments that are available in the training dataset (in our case Robust04) for learning the neural ranking model. These relevance assessments specify the relevance of query--document pairs in the data. 

We reconstruct the relevance assessments file using only the query--document pairs that were retrieved. The neural ranking model is then trained on the filtered query--document relevance file. 
When splitting the documents into passages the aforementioned design remains the same. 
This is explained in more detail in Section~\ref{sec:passageranking}.



\subsection{Re-Ranking Documents} \label{sec:reranking}



The second step of the ad-hoc retrieval pipeline is the re-ranking process. We use MatchZoo's (v2.2) implementation of the Deep Relevance Matching Model (DRMM) ~\cite{guo2016deep} to perform the neural re-ranking step. 
DRMM builds matching signals between query terms and documents using word embeddings and gives them as input to a feed-forward matching network that outputs a matching score for each query--document pair. 

\subsubsection{DRMM Model Architecture}

In more detail, DRMM takes as input the term vectors of a query $q = \{w_1^{(q)}, w_2^{(q)} , \ ... \ , w_M^{(q)}\}$ and a document $d = \{w_1^{(d)},w_2^{(d)}, \ ...\  , w_N^{(d)}\}$ where $w_i^{(q)}$ denotes the word embedding vector of the $i$-th query term and $w_j^{(d)}$ the word embedding vector of the $j$-th document term. These term vectors are used to build local interactions between each pair of terms. For each query term the local interactions are transformed into a fixed-length matching histogram. The matching histograms are the input of a feed-forward neural network which learns hierarchical matching patterns and produces a matching score for each query term. To generate the overall matching score of a query the scores for each term are aggregated with a term gating network. The matching score for each query-document pair is the dot product of the query term importance weights and the output vector of the feed-forward network.

\paragraph{Matching Histograms}

The first step in capturing local interactions between query terms and documents in DRMM is to measure the cosine similarity between the word vectors of every query term and every document term. The major drawback of this approach is the variance in the sizes of the local interactions due to the different lengths of queries in documents. In order to overcome this obstacle, DRMM introduced the matching histogram mapping which is a key element in its implementation. The basic idea behind matching histograms is the grouping of local interactions according to different levels of signal strength. The cosine similarity between the term vectors (a continuous value in the interval $[-1,1]$) is discretized into equally sized bins in an ascending order and each similarity score is assigned to the corresponding bin. The exact matching between a query and a document term (\textit{cosine similarity=1}) is treated as a separate bin. There are three types of matching histograms:

\begin{itemize}
    \item Count-based Histogram: Each bin contains the count of local interactions as the histogram value.
    \item Normalized Histogram: The count value in each bin is normalized by the total count of terms in the document.
    \item LogCount-based Histogram: The count value in each bin is logarithmized to reduce the range of histogram values. In our research, we only experimented with this type of matching histograms since previous work \cite{guo2016deep} shows they outperform the other ways of matching histogram mapping. 
\end{itemize}

\paragraph{feed-forward Matching Network}

Using the matching histograms as input, DRMM employs a feed-forward matching network to learn the hierarchical matching patterns between documents and query terms. The choice of a feed-forward network in the architecture of the model ignores the position of the terms and focuses on the strength of the signals. The  output of the network is a matching score for each query term.

\paragraph{Term Gating Network}
\label{sec:term_gating}
Another very important feature in the architecture of DRMM is the deployment of the term gating network. The purpose of this addition is the modeling of query term importance. Instead of just summing up the respective matching scores of each query term, the term gating network controls the contribution of each matching score to the final relevance score. To calculate this query term importance it produces an aggregation weight for each term using linear self-attention. Specifically, for each query $q_i$, its importance weight $g_i$ is returned by a softmax function:

\begin{equation}
    g_i = \frac{exp(w_g x_i^{(q)})}{\sum_{j=1}^{M}exp(w_g x_j^{(q)})} \ , \ \ i=1,... ,M
\label{eq:att_softmax}
\end{equation}

\noindent where $x_i^{(q)}$ is the $i$-th query term input. This input can be either the embedding vector of the corresponding term or its inverted document frequency. Consequently, $w_g$, which denotes the weight parameter of the gating network, is either a weight vector with the same dimensionality as the embedding vector or a scalar weight when the IDF value is chosen as the input.

\subsubsection{DRMM Model Training}

To train the deep relevance matching model we employ a pairwise ranking loss function, which is widely used in neural ranking and in ad-hoc information retrieval in general, called hinge loss.  Like in all pairwise loss functions, we compute hinge loss by taking the permutations of all document pairs. Given two matching scores $s(q,d^{+})$ and $s(q,d^{-})$ where $d^{+}$ is ranked higher than $d^{-}$ with respect to query $q$, hinge loss is defined as:

\begin{equation}
L(q,d^{+},d^{-} ; \Theta) = max(0,1 - s(q, d^{+}) + s(q, d^{-}))
\label{eq:hinge}
\end{equation}

\noindent with $\Theta$ as the weights used both in the feed-forward matching network and the term gating network. The model parameters are updated via standard backprogation using the Adadelta~\cite{zeiler2012adadelta} optimizer with mini-batches.

\subsection{Explainability} \label{sec:explainability}

\subsubsection{Query Term Importance}


Queries in web search engines are usually keyword-based and do not contain complex grammatical structures. Moreover, the compositional relation between the query terms can be often interpreted as logical `and', although not all terms should be of equal importance. Relying on this hypothesis DRMM handles queries as separate terms but takes into account the importance of each term by employing the term gating network (see Section~\ref{sec:term_gating}). 

We consider query term importance as an important aspect of search results interpretation. It is an actual real-time feedback to the user regarding the query they formulated in order to satisfy his information needs. The term weights could facilitate the search process and could teach the users how to search for what they are looking for, given a certain IR system.

To obtain the query term weights we use the term gating network's softmax function (see \autoref{eq:att_softmax}). This gating network can be alternatively described as the simplest form of an attention function \cite{vaswani2017attention}, a linear self-attention. Subsequently, the outputs $g_i$ of the gating network can be described as the outputs of an attention layer, also known as attention values. Since they are the output of a softmax function these probabilities sum to 1, which makes their visualization straightforward and interpretable.

\begin{figure}[t]
\centering
\includegraphics[width=\linewidth]{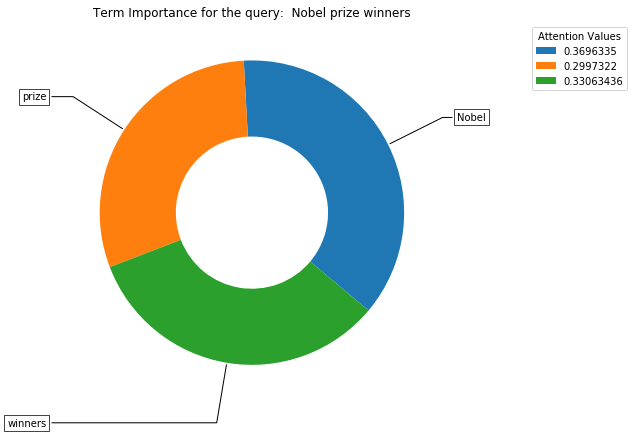}
\caption{A doughnut chart that depicts the importance of each term in the query "Nobel prize winners" based on their respective attention values from DRMM's term gating network}
\label{fig:nobel}
\end{figure}

\paragraph{Visualization of query term importance.} As example for our visualization we show the term importance for the query ``Nobel prize winners'' in \autoref{fig:nobel}. Given this query a relevant document is expected to be about winners of the Nobel prize. Intuitively the term "Nobel" is more important than the other two terms, in the sense that retrieved documents that do not refer to it are probably irrelevant and that documents describing other aspects of Nobel would be more relevant than documents about another prize for instance. As expected, the term \emph{Nobel} has the greatest attention probability among the other two terms while the term \emph{winners}, which is also very important, comes second.

\subsubsection{Passage ranking}\label{sec:passageranking}

For the explainability of the document text matching we decided to adopt a passage-level approach. We split each document in passages to be able to deduce the matching scores between the query and each passage, thereby allowing the selection of the best matching passage. Passages are non-overlapping and have a length of 100 tokens each. The small length of the passages improves the explainability of the model. Many documents in Robust04 are shorter than 200 tokens and could not be split into longer passages. Moreover smaller passages can be used as snippets for the search engine interface. 


Following DRMM's pipeline we build a matching histogram for each query--passage term pair and then give it as input to the model. Since passage-level labels are not available, in order to train the neural ranking model we have to assign a ground-truth relevance label to each passage-query pair. To that end, we transfer the document relevance label to each passage in the document as ground truth for learning the passage relevance.\footnote{This is a simple approach that might lead to some passages being mislabeled~\cite{akkalyoncu2019}, but for our current goal this suffices.}

After we have trained the model and assigned a matching score to each passage using DRMM, we assign the score of the passage with the highest match score to the document (`maxP'), following Dai et. al \cite{dai2019deeper} and use it to rank the documents. 

\begin{figure}[t]
\begin{subfigure}
\centering
\includegraphics[width=0.32\linewidth]{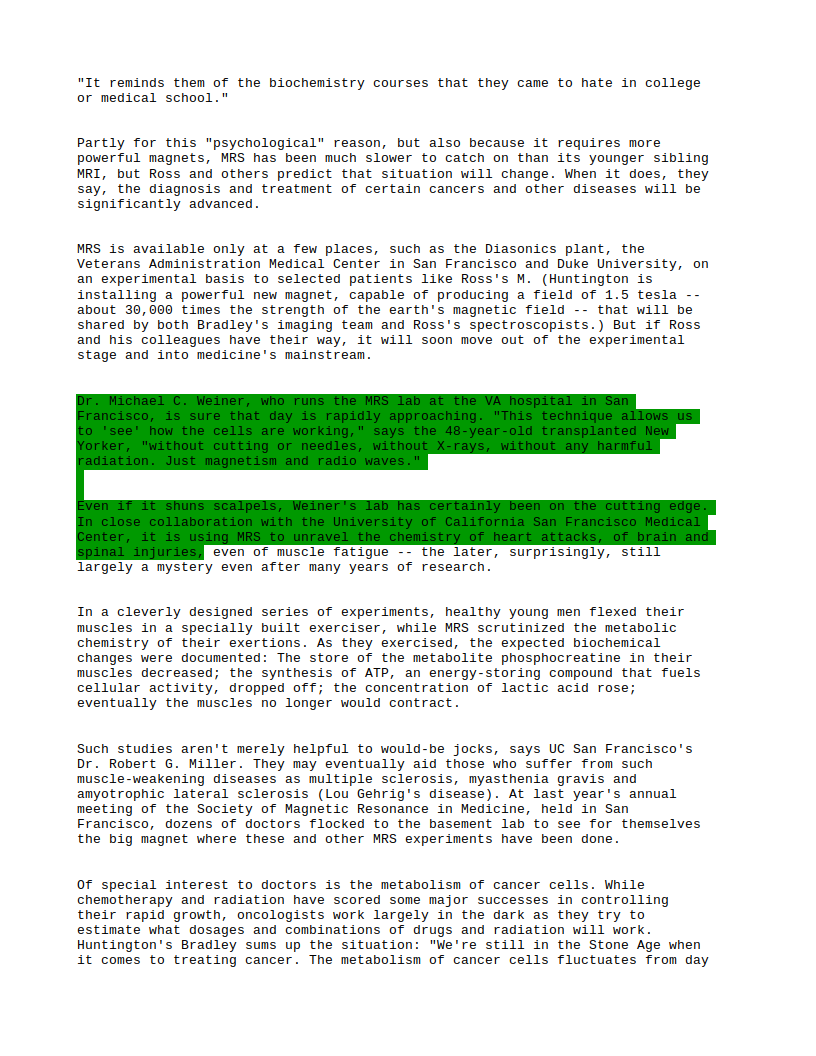}
\label{fig:doc1}
\end{subfigure}
\begin{subfigure}
\centering
\includegraphics[width=0.32\linewidth]{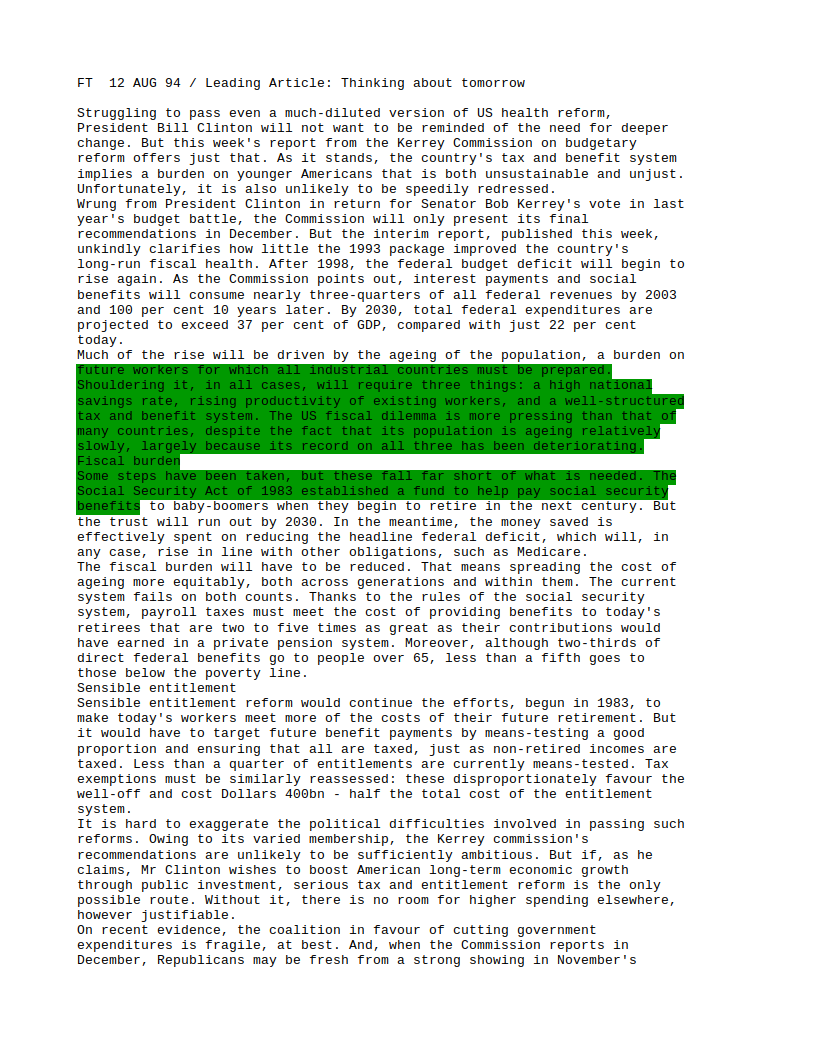}
\label{fig:doc2}
\end{subfigure}
\begin{subfigure}
\centering
\includegraphics[width=0.32\linewidth]{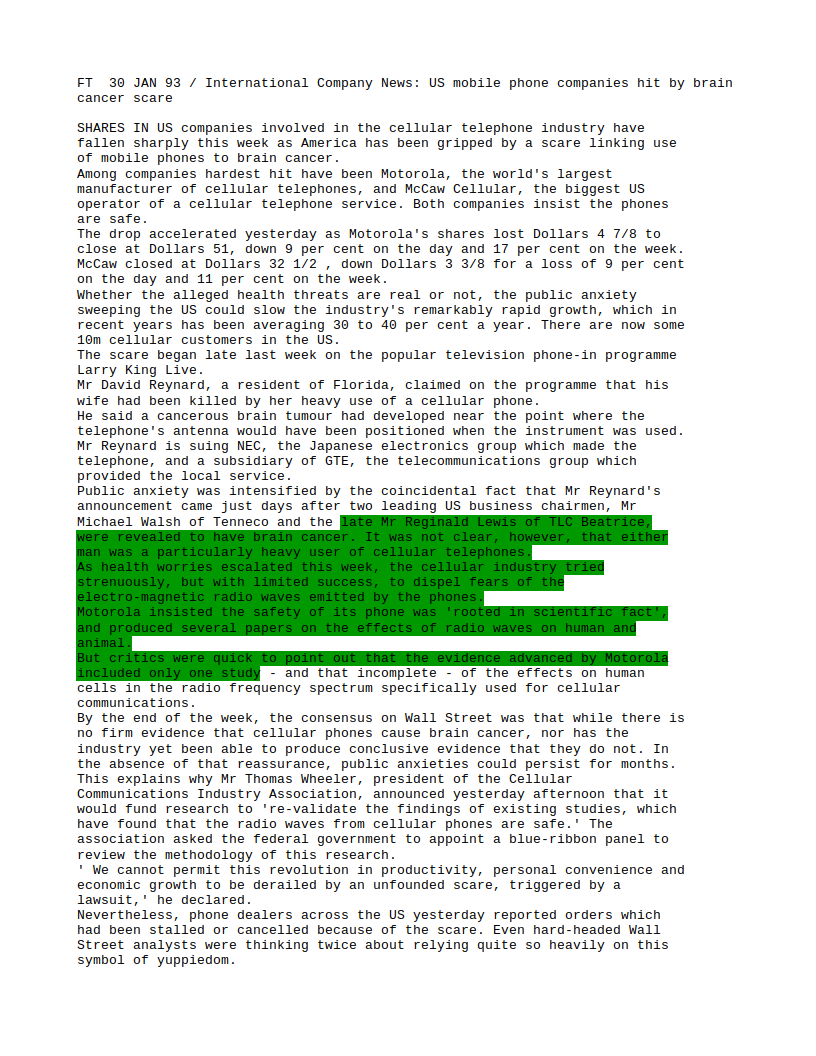}
\label{fig:doc3}

\end{subfigure}
\caption{Example thumbnails of retrieved documents with the most relevance passage highlighted.}
\label{fig:thumbs}
\end{figure}

\paragraph{Visualization of passage relevance.} The passage with the highest matching score from each query--document pair is considered to be the most relevant passage of the document and is shown to the user as document snippet in the search results interface. We also employ another type of visualization to enhance the explainability of the results. We show a thumbnail of the whole document to the user in which the most relevant passage is highlighted, as illustrated in Figure~\ref{fig:thumbs}. This allows the user to better judge the relevance of the document based on the position of the passage in the text and also gives information of where to search in the document for the most relevant passage. This can be especially helpful in long documents where the information needed by the user could be at the bottom of the document.

\begin{figure*}[t]
\begin{subfigure}
\centering
\frame{\includegraphics[height=5cm,trim={0 0 20cm 0},clip]{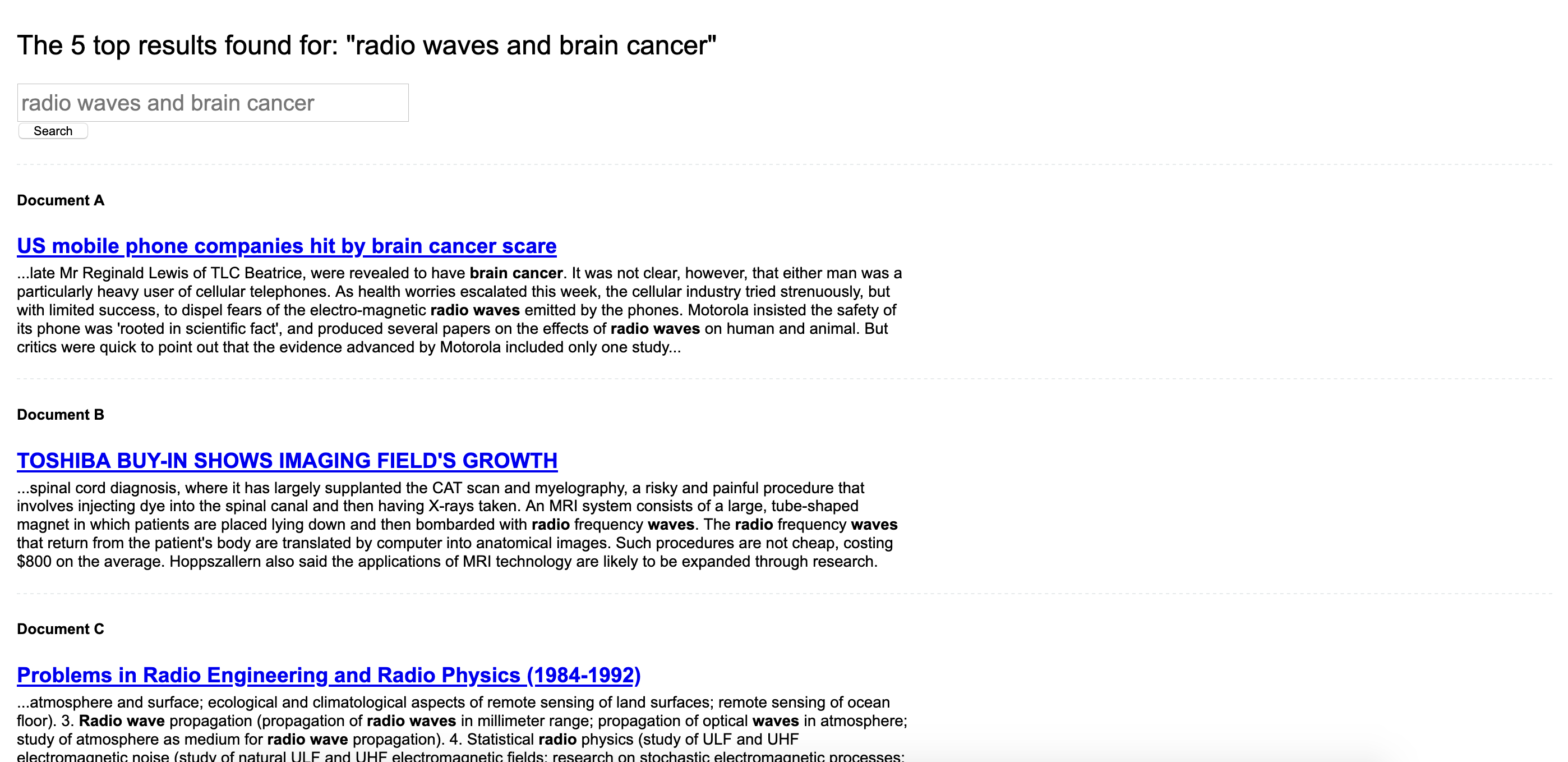}}
\label{fig:simple}
\end{subfigure}
\begin{subfigure}
\centering
\frame{\includegraphics[height=5cm]{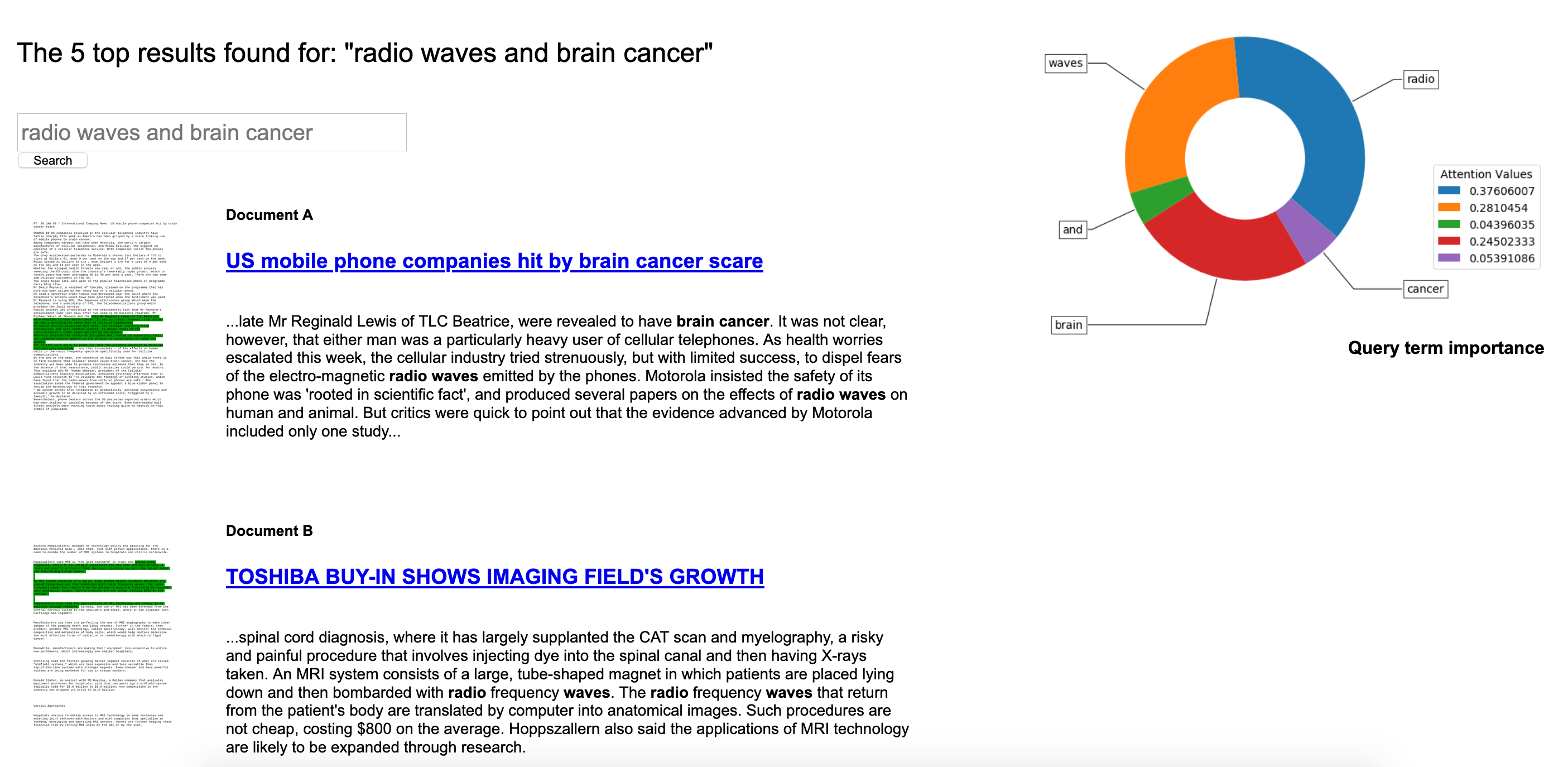}}
\label{fig:explainable}
\end{subfigure}
\caption{Left: Interface R (regular). Right: Interface E (explainable)}\label{fig:interfaces}
\end{figure*}

\section{Retrieval experiments} \label{sec:experiments}
\paragraph{Data.} We use the Robust04 dataset~\cite{voorhees2004overview} for the evaluation of our methods. Robust04 is a news corpus that contains 0.5M documents and 249 queries. The queries contain both a short title and a description. We only used the query titles in our experiments which are short (a few keywords) and therefore fit our the query term importance approach. In the initial retrieval step, we use $K=1000$ documents when retrieving complete documents and $K=100$ when splitting the documents into passages. We chose to reduce the number of initially retrieved documents in the passage approach because of computational limitations. The caveat of this approach is a drop in the performance of the initial retrieval model.


\paragraph{Neural re-ranking.} For the re-ranking step we split the dataset in 5 folds with 50 queries each, following Huston and Croft \cite{huston2014parameters}. We then conduct 5-fold cross-validation where we use 3 folds for training, 1 for validation and 1 for testing. For the DRMM implementation we use LogCount-based histograms with the bin size set to 30. In the feed-forward matching networks we used two hidden layers with 5 nodes each and \textit{tanh} as activation function. For the embedding layer we used Glove embeddings \cite{pennington2014glove} pretrained on 6B tokens with 300 dimensions.\footnote{https://nlp.stanford.edu/projects/glove/} The query term embeddings were used as the input of the term gating network.

\begin{table}[t]
\caption{Retrieval results on the Robust04 test data}
\begin{tabular}{l|c|c|c}
\hline
& MAP    & P@20   & nDCG@20 \\ 
 \hline
BM25 & 0.2531 & \textbf{0.3631} & \textbf{0.4240} \\ 
DRMM & 0.2662 & 0.2974 & 0.3706  \\
DRMM-maxP & \textbf{0.3172} & 0.2650 & 0.3177 \\ \hline
\end{tabular}
\label{tab:retrieval}
\end{table}

\paragraph{Results.} Table~\ref{tab:retrieval} shows the results of the retrieval experiments on the test set.
The first row in Table~\ref{tab:retrieval} are the results of Anserini's BM25 on the document level with $K=1000$ and acts as a strong baseline. The second row are the results of our DRMM implementation on the document level.\footnote{The results in terms of P@20 and nDCG@20 are lower than in the original paper~\cite{guo2016deep}, but our implementation is not identical to the original. Other papers have reported different results as well \cite{mcdonald2018deep}}
The third row is our own method for passage retrieval with evaluation on the document level, as explained in Section~\ref{sec:passageranking}. The results indicate that in terms of Precision@20 and nDCG@20 our model is less effective than BM25 and document-level DRMM, but it is a bit better in terms of MAP. Overall, we consider the quality of the model sufficient for evaluation of the explainable features developed.

\section{User evaluation} \label{sec:userstudy}
We created an explainable interface for the search engine result page including the query term relevance doughnut and the document thumbnails as described in Section~\ref{sec:explainability}. We compare our explainable interface to a regular result page interface in an online user study.

\subsection{Study design}
The study had a 2x2 within-subject design, with two groups of participants and two interfaces (regular and explainable). 

\paragraph{Queries and documents.} We hand-selected six queries from the Robust04 data to show to the participants in a result page together with the top-5 documents retrieved for the query. The queries were selected on the basis of three criteria: (1) that there is at least one relevant document in the top-5 retrieved documents; (2) that all top-5 retrieved documents have a title field;\footnote{Some documents in Robust04 do not have a title; these are much harder to assess on the result page. In addition, documents without titles are less representative for real live search engines} (3) that the query consists of at least three terms, to have an informative query doughnut chart. The selected queries are in Table~\ref{tab:queries}.

\begin{table}[t]
\caption{Selected queries for the user study. The relevant documents in the top 5 are indicated with letters, where A denotes the first-ranked document and E denotes the fifth-ranked document}
\begin{tabular}{p{1cm}|p{1cm}|l|p{1.5cm}}
\hline
our query id & Robust04 query id & query & relevant documents in top-5\\
\hline
1 & 310 & radio waves and brain cancer & A\\
2 & 318 & best retirement country & A,E\\
3 & 384 & space station moon & A\\
4 & 400 & amazon rain forest & C,D,E\\
5 & 613 & berlin wall disposal & B\\
6 & 615 & timber exports asia & B\\
\hline
\end{tabular}
\label{tab:queries}
\end{table}

For each query we retrieved and ranked documents from Robust04 using our retrieval and neural re-ranking method. We took the top-5 documents and for each document we selected the passage with the highest matching passage score as snippet. 

\paragraph{User interfaces.} We created two (static) search result page interfaces to compare, both shown in Figure~\ref{fig:interfaces}:
\begin{itemize}
    \item Interface R (regular) is a regular search engine result page with for each document the title and the most relevant passage as snippet. The query terms are marked with boldface in the snippet; 
    \item Interface E (explainable/experiment) is our experimental result page with for each document the title and most relevant passage as snippet. The query terms are marked with boldface in the snippet. In addition, the document thumbnail indicating the position of the snippet passage in the document is shown on the left of the snippet. On the top right of the page the doughnut chart for the query terms is shown.
\end{itemize}
Both interfaces have the same top-5 documents with the same snippets; only the visualization of the result page differs.

\paragraph{Participants.} 22 volunteers participated in the online study. Participants were randomly assigned to either group A or group B. Participants in group A saw query 1,3,5 in interface R and query 2,4,6 in interface E. Participants in group B saw query 2,4,6 in interface R and query 1,3,5 in interface E. The order of the queries was such that the two interfaces were shown to each user in alternating order, starting with interface R.

\subsection{Result page judgments}
Our evaluation criteria are based on the paper by Mi and Jiang on the interpretation of search result snippets \cite{mi2019understanding}, but as opposed to their work we evaluate the result page as a whole instead of the individual snippets. We ask our participants to judge the result pages on two criteria: explainability and assessability. According to Mi and Jang, assessability is ``the ability of a summary to explain search result relevance''. In our study, we generalize this to the result page as a whole: the ability of the result page to explain search result relevance. Participants were asked to respond to the following two judgment questions using a five point scale from 1 (Strongly Disagree) to 5 (Strongly Agree).

\begin{itemize}
    \item (\textbf{Explainability}) ``By looking at the result page, I can understand why the search engine returned these results for the shown query.''
    \item (\textbf{Assessability}) ``By looking at the result page, I can tell which results are useful without opening the links.''
\end{itemize}

In addition the participants were asked to assess the relevance of the 5 documents: ``By looking at the result page, can you tell which of the returned documents are relevant to the query?'' (checkboxes for document A--E, or `None of the above'). It was not possible for the users to click through to the full document; they only had the information on the search engine result page available.

\subsection{Results}
\begin{figure}[t]
\centering
\frame{\includegraphics[width=6cm]{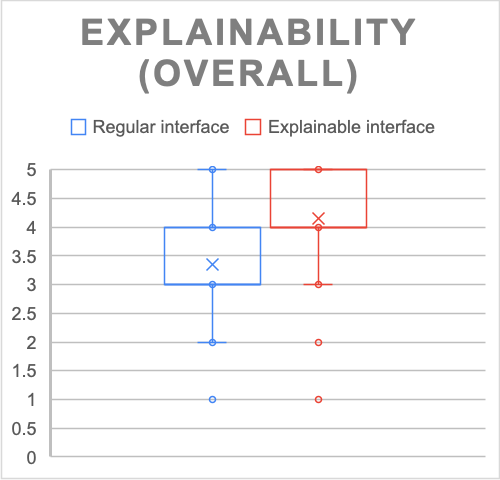}}
\caption{Dispersion of explainability scores for the two interfaces, over all queries and all participants (n=66 for each interface).}\label{fig:overall_expl}
\end{figure}

\begin{figure}[t]
\centering
\frame{\includegraphics[width=6cm]{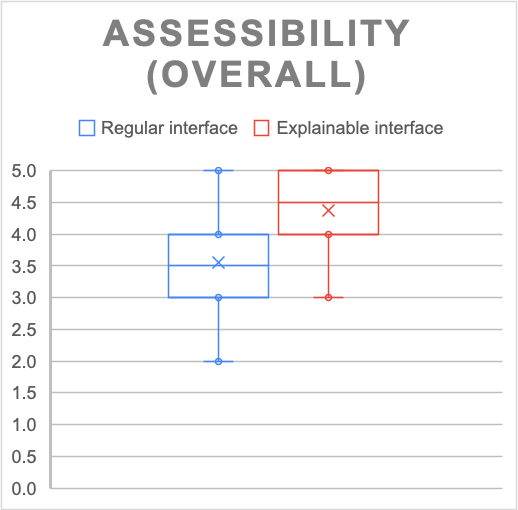}}
\caption{Dispersion of assessability scores for the two interfaces, over all queries and all participants (n=66 for each interface).}\label{fig:overall_assess}
\end{figure}

\begin{figure}[t]
\begin{subfigure}
\centering
\frame{\includegraphics[height=5cm]{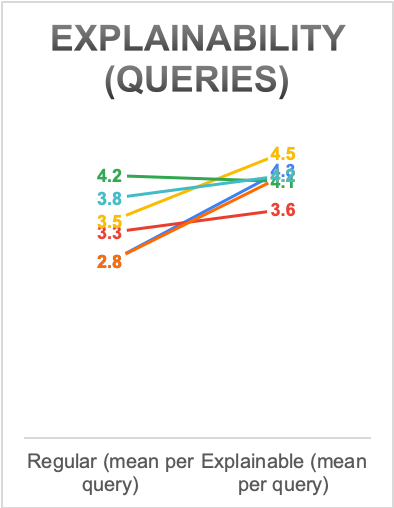}}
\label{fig:perquery_expl}
\end{subfigure}
\begin{subfigure}
\centering
\frame{\includegraphics[height=5cm]{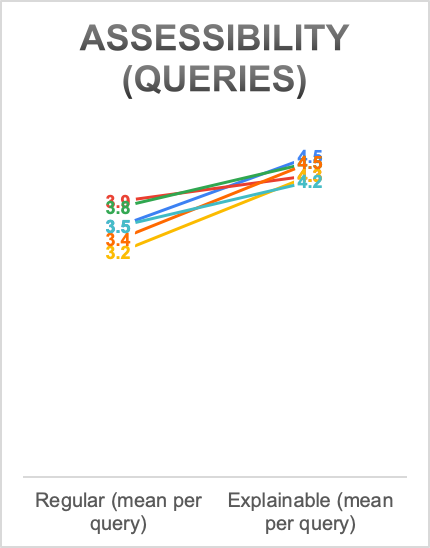}}
\end{subfigure}
\caption{Average explainability (left) and assessability (right) scores per query, in both interfaces. Each colored line denotes one query, with on the left side its mean score in the regular interface and on the right side its mean score in the explainable interface.}
\label{fig:perquery}
\end{figure}

\paragraph{Overall results for explainability and assessability.} The mean explainability score that the participants gave for the regular interface was 3.4; for the explainable interface this was higher: 4.2. Figure~\ref{fig:overall_expl} shows the dispersion of the explainability values over all queries and all participants. The mean assessability scores were 3.6 for the regular interface and 4.4 for the explainable interface. Figure~\ref{fig:overall_assess} shows the dispersion of the assessability values over all queries and all participants. 

\paragraph{Per-participant analysis.} Further analysis per participant with Wilcoxon Signed-ranks Tests (dependent samples with $n=22$) indicates that the scores for the two interfaces are significantly different (explainability: $W = 107, p = 0.001$ ; assessability: $W = 103.5, p = 0.001$) . The large majority of participants (18) judge the explainability of the explainable interface higher than the explainability of the regular interface. 

We conducted a three-way ANOVA to test the significance of the differences of the explainability and assessability scores between the two interfaces. 
The three factors (independent variables) of the analysis are the interface, the participant, and the query; the dependent variables are the explainability and assessability scores (we performed a three-way ANOVA for each of them separately). The analysis indicates that the \emph{explainability} scores are significantly different between the two interfaces ($F(1,21) = 30.7, p < 0.001$), but also between the participants ($F(1,21) = 3.4, p < 0.001$) and the queries ($F(1,21) = 3.3, p = 0.008$). The \emph{assessability} scores are significantly different between the participants ($F(1,21) = 3.1, p < 0.001$) and the interfaces ($F(1,1) = 39.6, p < 0.001$) but not between the queries ($p=0.24$). Moreover, the large F-Values of the interface variable show that there is a large variation between the two interfaces and we can conclude that the interface is the most significant factor for both assessability and explainability.

\paragraph{Per-query analysis.} Figure~\ref{fig:perquery} shows a breakout per query. For each of the six queries, the average explainability and assessability scores are plotted for the two interfaces. The left plot (for explainability) show quite some variation between the queries. This is in line with the ANOVA analysis, which indicated that the query is a significant factor in the variance of explainability scores. The query represented by the green line is the only query for which the explainable interface scores slightly lower (4.1) than the regular interface (4.2). This is query 4, ``amazon rain forest''. One characteristic of this query is that it has three relevant documents in the top-5, where the other queries had one and in one case two. This could have led to a relatively high explainability score in the regular interface (highest of all five queries in the regular interface). 

Assessability is judged higher for all queries in the explainable interface than in the regular interface. The relative difference between the interfaces (steepness of the lines) is similar between all six queries. This was also indicated by the ANOVA analysis, which showed that the query does is not a significant factor in the variance of the the assessability scores.

\paragraph{Relevance assessments.} We compared the participants' relevance assessments (based on the result page only) to the ground truth relevance assessments in the Robust04 data, and measured precision and recall per query, participant and interface type. The mean precision and recall scores per interface are in Table~\ref{tab:relass}. Although the scores seem higher for the explainable interface than for the regular interface, the deviation is large. Analysis on the participant level with the Wilcoxon Signed-ranks Test (dependent samples with $n=22$) indicates that the differences between the two interfaces in terms of precision and recall of the relevance assessments are not significant ($p=0.90$ and $p=0.38$, respectively).

\begin{table}[t]
\caption{Mean precision and recall scores (standard deviation between brackets) that the participants obtain by selecting snippets compared to the ground truth document relevance assessments in Robust04. }
\begin{tabular}{l|c|c}
\hline
 & Precision & Recall\\
\hline
Regular interface & 53\% (0.42) & 62\% (0.45)\\
Explainable interface & 59\% (0.40) & 75\% (0.41)\\
\hline
\end{tabular}
\label{tab:relass}
\end{table}

\section{Conclusions} \label{sec:conclusion}

\paragraph{RQ1. What is the ranking effectiveness of DRMM when selecting the most relevant passage of each document?}
We found that in terms of Precision@20 and nDCG@20 our passage-level ranking model is less effective than BM25 and document-level DRMM, but it is a bit better in terms of MAP. Overall, we consider the quality of the model to sufficient for evaluation of the explainable features developed.

\paragraph{RQ2. How do users judge the explainability and assessability of our explainable search engine result page compared to a regular result page?}
We found that users judge our proposed interface significantly more explainable and easier to assess than a regular search engine result page. This indicates that the explainability of the search engine result page leads to a better user experience. 

\paragraph{RQ3. How well can users select the relevant documents based on only the snippets on the result page, in the explainable interface compared to the regular interface? } We cannot prove that the users are better in selecting the relevant documents from the top-5; there is a large deviation in the precision and recall scores that the users obtain.

\vspace{0.5cm}
We conclude that the proposed explainable elements are promising as visualization for search engine users, based on their subjective experience. 

\paragraph{Future work.} We are interested in investigating the possibility of adding explainable elements to state-of-the-art neural ranking models for information retrieval. Neural ranking models like CEDR-DRMM \cite{macavaney2019cedr} and POSIT-DRMM \cite{mcdonald2018deep} are built on top of DRMM's architecture but improve its performance on ad-hoc retrieval tasks significantly. We believe that our methods can be modified accordingly in order to build explainable search engines that achieve performance comparable to the latest neural models. 

Furthermore, we would like to address explainable search in two research directions: professional search contexts, and personalized search contexts. We think that explainable search is particularly relevant for professional search contexts, where users are critical towards search results and have the need to be in control \cite{russell2018information,verberne2019first}. In these contexts, trust is even more important than in generic web search. In the case of personalized search, it is essential that the search engine is sufficiently transparent for two reasons: (1) to gain trust from the user that the personalization does not lead to loss in quality, even though other users get different results; (2) to be able to show the user which personal information is used in the result ranking, and make it explicit when documents are considered relevant because of matches to the users personal preferences.

\bibliography{explainable}


\begin{thebibliography}{24}


\ifx \showCODEN    \undefined \def \showCODEN     #1{\unskip}     \fi
\ifx \showDOI      \undefined \def \showDOI       #1{#1}\fi
\ifx \showISBNx    \undefined \def \showISBNx     #1{\unskip}     \fi
\ifx \showISBNxiii \undefined \def \showISBNxiii  #1{\unskip}     \fi
\ifx \showISSN     \undefined \def \showISSN      #1{\unskip}     \fi
\ifx \showLCCN     \undefined \def \showLCCN      #1{\unskip}     \fi
\ifx \shownote     \undefined \def \shownote      #1{#1}          \fi
\ifx \showarticletitle \undefined \def \showarticletitle #1{#1}   \fi
\ifx \showURL      \undefined \def \showURL       {\relax}        \fi
\providecommand\bibfield[2]{#2}
\providecommand\bibinfo[2]{#2}
\providecommand\natexlab[1]{#1}
\providecommand\showeprint[2][]{arXiv:#2}

\bibitem[\protect\citeauthoryear{Ai, Zhang, Bi, and Croft}{Ai
  et~al\mbox{.}}{2019}]%
        {ai2019explainable}
\bibfield{author}{\bibinfo{person}{Qingyao Ai}, \bibinfo{person}{Yongfeng
  Zhang}, \bibinfo{person}{Keping Bi}, {and} \bibinfo{person}{W~Bruce Croft}.}
  \bibinfo{year}{2019}\natexlab{}.
\newblock \showarticletitle{Explainable Product Search with a Dynamic Relation
  Embedding Model}.
\newblock \bibinfo{journal}{\emph{ACM Transactions on Information Systems
  (TOIS)}} \bibinfo{volume}{38}, \bibinfo{number}{1} (\bibinfo{year}{2019}),
  \bibinfo{pages}{1--29}.
\newblock


\bibitem[\protect\citeauthoryear{Akkalyoncu~Yilmaz, Yang, Zhang, and
  Lin}{Akkalyoncu~Yilmaz et~al\mbox{.}}{2019}]%
        {akkalyoncu2019}
\bibfield{author}{\bibinfo{person}{Zeynep Akkalyoncu~Yilmaz},
  \bibinfo{person}{Wei Yang}, \bibinfo{person}{Haotian Zhang}, {and}
  \bibinfo{person}{Jimmy Lin}.} \bibinfo{year}{2019}\natexlab{}.
\newblock \showarticletitle{Cross-Domain Modeling of Sentence-Level Evidence
  for Document Retrieval}. In \bibinfo{booktitle}{\emph{Proceedings of the 2019
  Conference on Empirical Methods in Natural Language Processing and the 9th
  International Joint Conference on Natural Language Processing
  (EMNLP-IJCNLP)}}. \bibinfo{publisher}{Association for Computational
  Linguistics}, \bibinfo{address}{Hong Kong, China},
  \bibinfo{pages}{3490--3496}.
\newblock
\urldef\tempurl%
\url{https://doi.org/10.18653/v1/D19-1352}
\showDOI{\tempurl}


\bibitem[\protect\citeauthoryear{Dai and Callan}{Dai and Callan}{2019}]%
        {dai2019deeper}
\bibfield{author}{\bibinfo{person}{Zhuyun Dai} {and} \bibinfo{person}{Jamie
  Callan}.} \bibinfo{year}{2019}\natexlab{}.
\newblock \showarticletitle{Deeper text understanding for IR with contextual
  neural language modeling}. In \bibinfo{booktitle}{\emph{Proceedings of the
  42nd International ACM SIGIR Conference on Research and Development in
  Information Retrieval}}. \bibinfo{pages}{985--988}.
\newblock


\bibitem[\protect\citeauthoryear{Fan, Pang, Hou, Guo, Lan, and Cheng}{Fan
  et~al\mbox{.}}{2017}]%
        {fan2017matchzoo}
\bibfield{author}{\bibinfo{person}{Yixing Fan}, \bibinfo{person}{Liang Pang},
  \bibinfo{person}{JianPeng Hou}, \bibinfo{person}{Jiafeng Guo},
  \bibinfo{person}{Yanyan Lan}, {and} \bibinfo{person}{Xueqi Cheng}.}
  \bibinfo{year}{2017}\natexlab{}.
\newblock \showarticletitle{Matchzoo: A toolkit for deep text matching}.
\newblock \bibinfo{journal}{\emph{arXiv preprint arXiv:1707.07270}}
  (\bibinfo{year}{2017}).
\newblock


\bibitem[\protect\citeauthoryear{Fang, Tao, and Zhai}{Fang
  et~al\mbox{.}}{2004}]%
        {fang2004formal}
\bibfield{author}{\bibinfo{person}{Hui Fang}, \bibinfo{person}{Tao Tao}, {and}
  \bibinfo{person}{ChengXiang Zhai}.} \bibinfo{year}{2004}\natexlab{}.
\newblock \showarticletitle{A formal study of information retrieval
  heuristics}. In \bibinfo{booktitle}{\emph{Proceedings of the 27th annual
  international ACM SIGIR conference on Research and development in information
  retrieval}}. \bibinfo{pages}{49--56}.
\newblock


\bibitem[\protect\citeauthoryear{Fernando, Singh, and Anand}{Fernando
  et~al\mbox{.}}{2019}]%
        {fernando2019study}
\bibfield{author}{\bibinfo{person}{Zeon~Trevor Fernando},
  \bibinfo{person}{Jaspreet Singh}, {and} \bibinfo{person}{Avishek Anand}.}
  \bibinfo{year}{2019}\natexlab{}.
\newblock \showarticletitle{A study on the Interpretability of Neural Retrieval
  Models using DeepSHAP}. In \bibinfo{booktitle}{\emph{Proceedings of the 42nd
  International ACM SIGIR Conference on Research and Development in Information
  Retrieval}}. \bibinfo{pages}{1005--1008}.
\newblock


\bibitem[\protect\citeauthoryear{Guo, Fan, Ai, and Croft}{Guo
  et~al\mbox{.}}{2016}]%
        {guo2016deep}
\bibfield{author}{\bibinfo{person}{Jiafeng Guo}, \bibinfo{person}{Yixing Fan},
  \bibinfo{person}{Qingyao Ai}, {and} \bibinfo{person}{W~Bruce Croft}.}
  \bibinfo{year}{2016}\natexlab{}.
\newblock \showarticletitle{A deep relevance matching model for ad-hoc
  retrieval}. In \bibinfo{booktitle}{\emph{Proceedings of the 25th ACM
  International on Conference on Information and Knowledge Management}}.
  \bibinfo{pages}{55--64}.
\newblock


\bibitem[\protect\citeauthoryear{Huston and Croft}{Huston and Croft}{2014}]%
        {huston2014parameters}
\bibfield{author}{\bibinfo{person}{Samuel Huston} {and}
  \bibinfo{person}{W~Bruce Croft}.} \bibinfo{year}{2014}\natexlab{}.
\newblock \bibinfo{booktitle}{\emph{Parameters learned in the comparison of
  retrieval models using term dependencies}}.
\newblock \bibinfo{type}{{T}echnical {R}eport}.
\newblock


\bibitem[\protect\citeauthoryear{MacAvaney, Yates, Cohan, and
  Goharian}{MacAvaney et~al\mbox{.}}{2019}]%
        {macavaney2019cedr}
\bibfield{author}{\bibinfo{person}{Sean MacAvaney}, \bibinfo{person}{Andrew
  Yates}, \bibinfo{person}{Arman Cohan}, {and} \bibinfo{person}{Nazli
  Goharian}.} \bibinfo{year}{2019}\natexlab{}.
\newblock \showarticletitle{CEDR: Contextualized embeddings for document
  ranking}. In \bibinfo{booktitle}{\emph{Proceedings of the 42nd International
  ACM SIGIR Conference on Research and Development in Information Retrieval}}.
  \bibinfo{pages}{1101--1104}.
\newblock


\bibitem[\protect\citeauthoryear{McDonald, Brokos, and
  Androutsopoulos}{McDonald et~al\mbox{.}}{2018}]%
        {mcdonald2018deep}
\bibfield{author}{\bibinfo{person}{Ryan McDonald},
  \bibinfo{person}{Georgios-Ioannis Brokos}, {and} \bibinfo{person}{Ion
  Androutsopoulos}.} \bibinfo{year}{2018}\natexlab{}.
\newblock \showarticletitle{Deep relevance ranking using enhanced
  document-query interactions}.
\newblock \bibinfo{journal}{\emph{arXiv preprint arXiv:1809.01682}}
  (\bibinfo{year}{2018}).
\newblock


\bibitem[\protect\citeauthoryear{Mi and Jiang}{Mi and Jiang}{2019}]%
        {mi2019understanding}
\bibfield{author}{\bibinfo{person}{Siyu Mi} {and} \bibinfo{person}{Jiepu
  Jiang}.} \bibinfo{year}{2019}\natexlab{}.
\newblock \showarticletitle{Understanding the Interpretability of Search Result
  Summaries}. In \bibinfo{booktitle}{\emph{Proceedings of the 42nd
  International ACM SIGIR Conference on Research and Development in Information
  Retrieval}}. \bibinfo{pages}{989--992}.
\newblock


\bibitem[\protect\citeauthoryear{Pennington, Socher, and Manning}{Pennington
  et~al\mbox{.}}{2014}]%
        {pennington2014glove}
\bibfield{author}{\bibinfo{person}{Jeffrey Pennington},
  \bibinfo{person}{Richard Socher}, {and} \bibinfo{person}{Christopher~D
  Manning}.} \bibinfo{year}{2014}\natexlab{}.
\newblock \showarticletitle{Glove: Global vectors for word representation}. In
  \bibinfo{booktitle}{\emph{Proceedings of the 2014 conference on empirical
  methods in natural language processing (EMNLP)}}.
  \bibinfo{pages}{1532--1543}.
\newblock


\bibitem[\protect\citeauthoryear{Rennings, Moraes, and Hauff}{Rennings
  et~al\mbox{.}}{2019}]%
        {rennings2019axiomatic}
\bibfield{author}{\bibinfo{person}{Dani{\"e}l Rennings},
  \bibinfo{person}{Felipe Moraes}, {and} \bibinfo{person}{Claudia Hauff}.}
  \bibinfo{year}{2019}\natexlab{}.
\newblock \showarticletitle{An axiomatic approach to diagnosing neural ir
  models}. In \bibinfo{booktitle}{\emph{European Conference on Information
  Retrieval}}. Springer, \bibinfo{pages}{489--503}.
\newblock


\bibitem[\protect\citeauthoryear{Ribeiro, Singh, and Guestrin}{Ribeiro
  et~al\mbox{.}}{2016}]%
        {ribeiro2016model}
\bibfield{author}{\bibinfo{person}{Marco~Tulio Ribeiro},
  \bibinfo{person}{Sameer Singh}, {and} \bibinfo{person}{Carlos Guestrin}.}
  \bibinfo{year}{2016}\natexlab{}.
\newblock \showarticletitle{Model-agnostic interpretability of machine
  learning}.
\newblock \bibinfo{journal}{\emph{arXiv preprint arXiv:1606.05386}}
  (\bibinfo{year}{2016}).
\newblock


\bibitem[\protect\citeauthoryear{Robertson and Walker}{Robertson and
  Walker}{1994}]%
        {BM25robertson1994some}
\bibfield{author}{\bibinfo{person}{Stephen~E Robertson} {and}
  \bibinfo{person}{Steve Walker}.} \bibinfo{year}{1994}\natexlab{}.
\newblock \showarticletitle{Some simple effective approximations to the
  2-poisson model for probabilistic weighted retrieval}. In
  \bibinfo{booktitle}{\emph{SIGIR’94}}. Springer, \bibinfo{pages}{232--241}.
\newblock


\bibitem[\protect\citeauthoryear{Russell-Rose, Chamberlain, and
  Azzopardi}{Russell-Rose et~al\mbox{.}}{2018}]%
        {russell2018information}
\bibfield{author}{\bibinfo{person}{Tony Russell-Rose}, \bibinfo{person}{Jon
  Chamberlain}, {and} \bibinfo{person}{Leif Azzopardi}.}
  \bibinfo{year}{2018}\natexlab{}.
\newblock \showarticletitle{Information retrieval in the workplace: A
  comparison of professional search practices}.
\newblock \bibinfo{journal}{\emph{Information Processing \& Management}}
  \bibinfo{volume}{54}, \bibinfo{number}{6} (\bibinfo{year}{2018}),
  \bibinfo{pages}{1042--1057}.
\newblock


\bibitem[\protect\citeauthoryear{Singh and Anand}{Singh and Anand}{2019}]%
        {singh2019exs}
\bibfield{author}{\bibinfo{person}{Jaspreet Singh} {and}
  \bibinfo{person}{Avishek Anand}.} \bibinfo{year}{2019}\natexlab{}.
\newblock \showarticletitle{EXS: Explainable search using local model agnostic
  interpretability}. In \bibinfo{booktitle}{\emph{Proceedings of the Twelfth
  ACM International Conference on Web Search and Data Mining}}.
  \bibinfo{pages}{770--773}.
\newblock


\bibitem[\protect\citeauthoryear{Vaswani, Shazeer, Parmar, Uszkoreit, Jones,
  Gomez, Kaiser, and Polosukhin}{Vaswani et~al\mbox{.}}{2017}]%
        {vaswani2017attention}
\bibfield{author}{\bibinfo{person}{Ashish Vaswani}, \bibinfo{person}{Noam
  Shazeer}, \bibinfo{person}{Niki Parmar}, \bibinfo{person}{Jakob Uszkoreit},
  \bibinfo{person}{Llion Jones}, \bibinfo{person}{Aidan~N Gomez},
  \bibinfo{person}{{\L}ukasz Kaiser}, {and} \bibinfo{person}{Illia
  Polosukhin}.} \bibinfo{year}{2017}\natexlab{}.
\newblock \showarticletitle{Attention is all you need}. In
  \bibinfo{booktitle}{\emph{Advances in neural information processing
  systems}}. \bibinfo{pages}{5998--6008}.
\newblock


\bibitem[\protect\citeauthoryear{Verberne, He, Kruschwitz, Wiggers, Larsen,
  Russell-Rose, and de~Vries}{Verberne et~al\mbox{.}}{2019}]%
        {verberne2019first}
\bibfield{author}{\bibinfo{person}{Suzan Verberne}, \bibinfo{person}{Jiyin He},
  \bibinfo{person}{Udo Kruschwitz}, \bibinfo{person}{Gineke Wiggers},
  \bibinfo{person}{Birger Larsen}, \bibinfo{person}{Tony Russell-Rose}, {and}
  \bibinfo{person}{Arjen~P de Vries}.} \bibinfo{year}{2019}\natexlab{}.
\newblock \showarticletitle{First International Workshop on Professional
  Search}. In \bibinfo{booktitle}{\emph{ACM SIGIR Forum}},
  Vol.~\bibinfo{volume}{52}. ACM New York, NY, USA, \bibinfo{pages}{153--162}.
\newblock


\bibitem[\protect\citeauthoryear{Voorhees}{Voorhees}{2004}]%
        {voorhees2004overview}
\bibfield{author}{\bibinfo{person}{Ellen~M Voorhees}.}
  \bibinfo{year}{2004}\natexlab{}.
\newblock \showarticletitle{Overview of TREC 2004.}. In
  \bibinfo{booktitle}{\emph{Trec}}.
\newblock


\bibitem[\protect\citeauthoryear{Yang, Fang, and Lin}{Yang
  et~al\mbox{.}}{2017}]%
        {yang2017anserini}
\bibfield{author}{\bibinfo{person}{Peilin Yang}, \bibinfo{person}{Hui Fang},
  {and} \bibinfo{person}{Jimmy Lin}.} \bibinfo{year}{2017}\natexlab{}.
\newblock \showarticletitle{Anserini: Enabling the use of Lucene for
  information retrieval research}. In \bibinfo{booktitle}{\emph{Proceedings of
  the 40th International ACM SIGIR Conference on Research and Development in
  Information Retrieval}}. \bibinfo{pages}{1253--1256}.
\newblock


\bibitem[\protect\citeauthoryear{Zeiler}{Zeiler}{2012}]%
        {zeiler2012adadelta}
\bibfield{author}{\bibinfo{person}{Matthew~D Zeiler}.}
  \bibinfo{year}{2012}\natexlab{}.
\newblock \showarticletitle{Adadelta: an adaptive learning rate method}.
\newblock \bibinfo{journal}{\emph{arXiv preprint arXiv:1212.5701}}
  (\bibinfo{year}{2012}).
\newblock


\bibitem[\protect\citeauthoryear{Zhang, Mao, and Ai}{Zhang
  et~al\mbox{.}}{2019a}]%
        {zhang2019sigir}
\bibfield{author}{\bibinfo{person}{Yongfeng Zhang}, \bibinfo{person}{Jiaxin
  Mao}, {and} \bibinfo{person}{Qingyao Ai}.} \bibinfo{year}{2019}\natexlab{a}.
\newblock \showarticletitle{{SIGIR 2019 tutorial on explainable recommendation
  and search}}. In \bibinfo{booktitle}{\emph{Proceedings of the 42nd
  International ACM SIGIR Conference on Research and Development in Information
  Retrieval}}. \bibinfo{pages}{1417--1418}.
\newblock


\bibitem[\protect\citeauthoryear{Zhang, Zhang, Zhang, and Shah}{Zhang
  et~al\mbox{.}}{2019b}]%
        {zhang2019ears}
\bibfield{author}{\bibinfo{person}{Yongfeng Zhang}, \bibinfo{person}{Yi Zhang},
  \bibinfo{person}{Min Zhang}, {and} \bibinfo{person}{Chirag Shah}.}
  \bibinfo{year}{2019}\natexlab{b}.
\newblock \showarticletitle{EARS 2019: The 2nd international workshop on
  explainable recommendation and search}. In
  \bibinfo{booktitle}{\emph{Proceedings of the 42nd International ACM SIGIR
  Conference on Research and Development in Information Retrieval}}.
  \bibinfo{pages}{1438--1440}.
\newblock


\end{thebibliography}
\bibliographystyle{ACM-Reference-Format}

\end{document}